# On a Search for Hidden Photon CDM by a Multi-Cathode Counter

A.V.Kopylov[1], I.V.Orekhov and V.V.Petukhov

*Institute for Nuclear Research of Russian Academy of Sciences,*
*117312, Prospect of 60th October Revolution 7A, Russia*
*E-mail*: beril@inr.ru

ABSTRACT: We report on a new technique of a Multi-Cathode Counter (MCC) developed to search for hidden photon (HP) cold dark matter (CDM) with a mass from 5 to 500 eV. The method is suggested in the assumption that HP-photon mixing causes emission of single electrons from a metal cathode if the mass of hidden photon $m_{\gamma'}$ is greater than a work function of the metal $\varphi_W$. The measured effect from HP should be dependent on $\varphi_W$ and on the structure of electronic shells of the metal used as a cathode. Potentially this can be used for a verification of the results obtained. Some preliminary results for the upper limit for mixing parameter $\chi$ have been obtained for HP with a mass from 5 eV to 10 keV as a pure illustration of the potential of this technique. The efforts are continued to refine the procedure of data treatment and to improve the work of MCC. A new detector with a more developed design is under construction.

## 1. Introduction

The present data on the structure formation in the Universe indicate that most Dark Matter (DM) is "cold" i.e. should be non-relativistic. Neutrino in a Hot Dark Matter concept can be envisaged only in combination with Cold Dark Matter (CDM). Now the most attractive DM candidates appear to be Weakly Interactive Massive Particles (WIMP). Great progress in this field of research is outlined in [1]. However there are other alternatives, among them axion and axion-like particles (ALP) which is probably a next, most promising field. The efforts to discover axion are described in details in [2]. Another interesting opportunity is a hidden photon which is a light extra U(1) gauge bozon. As it was suggested in [3, 4] hidden photons (HP) may be observed in experiment through a kinetic mixing term $(\chi/2)F_{\mu\nu}X^{\mu\nu}$ with the ordinary photons, where $\chi$ is a parameter quantifying the kinetic mixing. Here $F_{\mu\nu}$ is the field stress of the ordinary electromagnetic field $A^\mu$ and $X^{\mu\nu}$ is the field stress of the HP field $X^\mu$.

Recently the eV mass range of HP was investigated with a dish antenna [5], a novel method proposed in [6]. The idea is to detect electromagnetic wave which is emitted by the oscillation of electrons of the antenna's surface under tiny ordinary electromagnetic field $A^\mu$ induced by HP. A dark matter solution for HP with a mass $m_{\gamma'}$ reads [6]

$$\left.\begin{pmatrix} \mathbf{A} \\ \mathbf{X} \end{pmatrix}\right|_{DM} = \mathbf{X}_{DM} \begin{pmatrix} -\chi \\ 1 \end{pmatrix} \exp(-i\omega t) \qquad (1)$$

i.e. has a spatially constant mode k = 0, oscillating with frequency $\omega = m_{\gamma'}$. This method works well only if the reflectance of antenna is high what is observed for $\omega$ < 5 eV. In the work [5] using an optical mirror and a photon-counting PMT at the point of convergence of the photons emitted from mirror the upper limit of 6x10$^{-12}$ was obtained for a mixing parameter χ for the hidden photon mass $m_{\gamma'}$= 3.1 ± 1.2 eV. This work was a first measurement of χ within this range of $m_{\gamma'}$ using a dish antenna. Our work was the search for hidden photons for the upper range of $m_{\gamma'}$ using a gaseous proportional counter as a detector of electrons emitted from a metal cathode by hidden photons. This constitutes the novelty of our method.

---
[1] Corresponding author.

## 2. A Method and experimental apparatus

The principal difference of our approach from [6] is that here we focus on shorter wavelengths, i.e. higher masses of HPs for which the reflectance of antenna is low. We make an assumption that in this case a HP-photon conversion will cause emission of single electrons from the surface of antenna similar to what one observes when metal is exposed to UV radiation. To register this conversion the detector should be highly sensitive to single electrons emitted from metal. Here we would like to draw attention to the fact that the detector in this case is sensitive exclusively to HP-photon conversion, not to photon-HP conversion. For the latter one this method does not work. We assume here that probability for the electron to be emitted after a hidden photon gets converted into ordinary electric field in a metal cathode would be equal to quantum efficiency η for a given metal to emit electron after absorption of a real UV photon of energy ω = $m_{\gamma'}$. The details of how electrons are created after photons get absorbed by a metal photocathode are described in many papers, see, for example [7]. The obvious difference between these two processes is that UV photons are strongly absorbed by the metal while HPs move freely through the metal, but this lead only to some underestimation of the effect from HP-photon conversion. We did not take this into account. We neither took into account the effects of surface roughness of the cathode. We consider this to be small corrections and leave it for our further study. To make a practical implementation of this idea a special technique of a Multi-Cathode Counter (MCC) has been developed and some very preliminary data as a pure illustration of the potential of this technique in the search for hidden photons has been obtained.

To count electrons emitted from a metal cathode we used a gaseous proportional counter filled by argon – methane (10%) mixture at 0.2 MPa. To detect single electrons the counter should have high (≥ 10$^5$) coefficient of gas amplification. The general view of the counter is presented on Fig.1 and the electronic scheme on Figure 2. Present design of MCC first described in [8] is a further development of the work with the aim to make an apparatus to register neutrino - nucleus coherent scattering [9, 10]. The cathode of the counter is 194 mm in diameter and 400 mm in length. It has relatively large (≈ 0.2 m$^2$) surface which acts in this experiment as "antenna" for HP but instead of reflecting electromagnetic waves it emits single electrons. The counter has a central anode wire of 20 μm and 4 cathodes, 3 of them are composed of an array of 50 μm nichrome wires tensed with a pitch of a few mm around anode one after another, and a fourth one, more distant from anode, is a cathode made of copper.

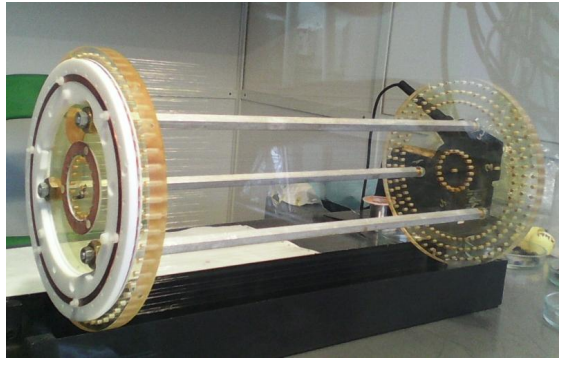

Figure 1. The central part of the counter.

The apparatus is counting electrons emitted from the walls of a cathode at short wavelengths $\omega = m_{\gamma'} \approx 5 - 500$ eV. The diameter of the first cathode $D_1$ is 40 mm to ensure high ($\geq 10^5$) coefficient of gas amplification in the central section of the counter. Three different configurations of the same counter are used to measure the count rate of single electrons. In the first configuration electrons, emitted from copper drift freely to the central section with high gas amplification. The highest negative potential is applied in this configuration to the copper cathode. The rate $R_1$ measured in this configuration

$$R_1 = R_{Cu} + R_{sp} \qquad (2)$$

Here $R_{Cu}$ is the count rate from single electrons emitted from copper and $R_{sp}$ – the rate from spurious pulses generated in the volume limited by a diameter 194 mm. What is the origin of spurious pulses, this question will be the subject of our further study. Here we assume that they are produced by "hot" spots on the surface of metal and are the effect of microstructure of a metal. The micro protrusions (spearheads) on the surface of wires may generate single electrons in strong electric fields. The spots with impurities of other metals, especially alkaline, may generate emission of single electrons etc. Only the future work can show how successful can be this approach.

In the second configuration the highest negative potential is applied to the third cathode $D_3 = 180$ mm. In this configuration the count rate

$$R_2 = 0.11 \cdot R_{Cu} + R_{sp} \qquad (3)$$

The factor 0.11 was obtained by calibration of the counter in 1st and 2nd configurations by UV source of the same intensity. Here the key point is that in 1st and 2nd configurations the counter has approximately the same geometry and the same wires. So if "hot" spots are on the surface of wires, the difference of the count rates in 1st and 2nd configurations will contain only 89% of the count rate of single electrons emitted from copper cathode. Because the geometry of the counter in 1st configuration is very similar to its geometry in 2nd configuration the same reasoning is valid for spurious pulses generated by a leakage current through dielectric used in the construction of the counter.

In the third configuration the highest negative potential is applied to a second cathode $D_2 = 140$ mm. The rate $R_3$ measured in this configuration is determined by spurious pulses generated within a volume limited by smaller diameter 140 mm with different number of wires and different isolators, so:

$$R_3 = r_{sp} \qquad (4)$$

In experiment the rate $R_3$ turned out to be approximately 3 times smaller than the rates $R_1$ and $R_2$ (which are very close) not in contradiction with our explanation of the origin of the spurious pulses. All this explains why as a measure of the effect from HPs we use the count rate

$$R_{MCC} = (R_1 - R_2)/0.89 = R_{Cu} \qquad (5)$$

The measurements were performed in each configuration by switching the counter consecutively in three different configurations, by calibration of the counter and by measuring the rates $R_1$, $R_2$ and $R_3$. Then from a number of measured points the average rates $\overline{R}_1$, $\overline{R}_2$ were found and from here - the average rate $\overline{R}_{MCC}$. Then from the scattering of the experimental points the uncertainties were calculated for each rate and, finally, for $R_{MCC}$. The count rate $R_3$ has been used to monitor the counting process to exclude some possible interference by external sources of electromagnetic disturbances.

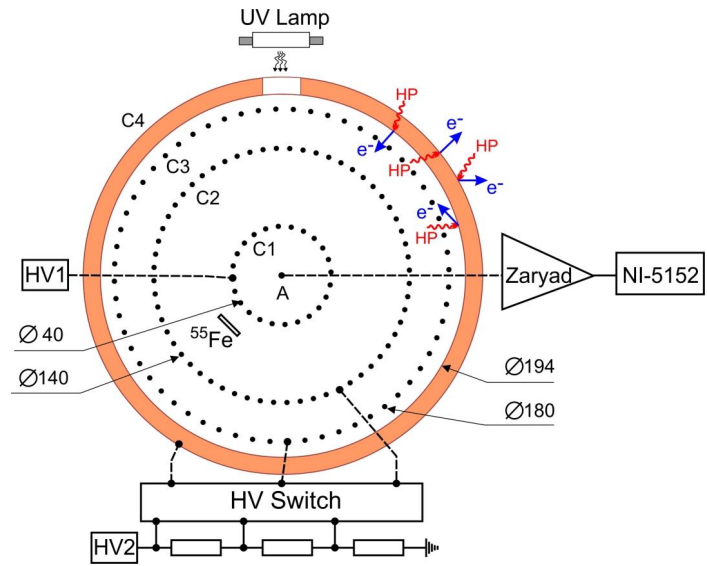

Figure 2. A simplified electronic scheme of a multi-cathode counter (MCC).
A – anode, C1 – C4 – cathodes, "Zaryad" – charge sensitive preamplifier.

## 3. Energy calibration and analysis

The calibration of the counter has been conducted by $^{55}$Fe source and by UV light of the mercury lamp. The calibration by $^{55}$Fe source was used to determine at what high voltages HV1 and HV2 the counter was working in a mode of limited proportionality with high ($>10^5$) gas amplification. Here we followed the standard technique described in many papers, for example in [14]. It was also described in our earlier papers [9, 10]. High voltage at first cathode was 2060 V and the ones from the voltage divider have been used for all three configurations such as to ensure the amplitude of the pulse corresponding to peak 5.9 keV from K-line of $^{55}$Mn, which is eradiated as a result of K-electron capture by $^{55}$Fe, on the output of charge sensitive preamplifier to be at the level 1400 mV what corresponds to a gas amplification $A \approx 10^5$. Figure 3 shows the calibration spectrum.

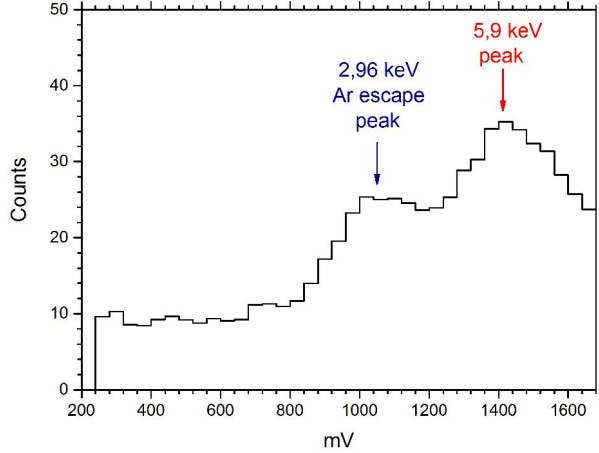

Figure 3. The calibration spectrum from $^{55}$Fe source introduced directly in a working volume of the counter facing first cathode.

The amplitude of the escape peak of argon at the energy 2.96 keV (5.9 keV minus 2.94 keV of K X-ray of argon escaping the detection region) was shifted from the due position in case of a linear response at about 700 mV to the one at approximately 1000 mV. This nonlinearity in the spectrum indicates that the counter was working in the regime of limited proportionality. From the approximation done through three points: zero and two peaks: 5.9 keV at 1400 mV and 2.96 keV at 1000 mV one can find that at energies less than 100 eV a gas amplification $A \approx 1.8 \cdot 10^5$ and the conversion factor is $\approx 2.3$ eV/mV. It takes approximately 27 eV to create one electron-ion pair in argon. It means that single electron pulses should be observed in the region below 50 mV. The calibration by UV photons demonstrates that this is really so. The $^{55}$Fe source has been removed and the internal walls of the counter were irradiated by UV light from a mercury lamp placed outside through a window made of melted silica. The physics and techniques of single electron counting were described in details in many devoted articles; see, for example [11 - 13]. Figure 4 shows the single electron spectra obtained in measurements in 1$^{st}$ and 2$^{nd}$ configurations.

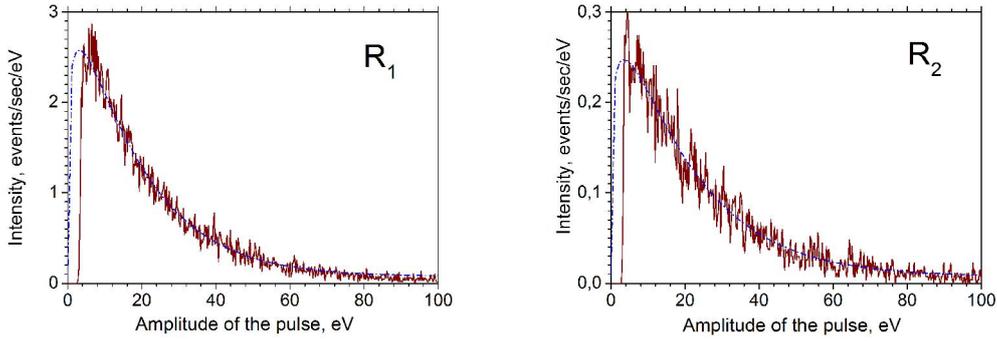

Figure 4. The single electron spectra obtained in measurements in
1st ($R_1$) and 2nd ($R_2$) configurations at the same flux of UV photons.
The dashed curves indicate the Polya distribution.

Comparing the count rates $R_1$ and $R_2$ presented on Fig.4 one can see that in the 2nd configuration the count rate of single electrons was about 0.11 of the rate measured in the first one. It proves that in the 2nd configuration the electrons emitted from the walls of the counter were really rejected back by the 3rd cathode. It means that the counter in the 2nd configuration can be used for measurement of the background count rate. To compare the gains in different configurations we used the same procedure as a one described in [15]; the inverse indexes of exponents were used as a measure of the gain of the counter. For the pulses above a threshold 7 eV the deviation from exponential distribution was small and could be neglected. The inverse indexes of exponent for all three configurations were in the range 19.3 ± 1.2 eV. The counting efficiency for the interval from 7 eV till 70 eV was found to be 76 ± 5 %. This number has been obtained by using Polya distribution [16]

$$P(A) = \left(\frac{A(1+\theta)}{\bar{A}}\right)^{\theta} \exp\left(\frac{-A(1+\theta)}{\bar{A}}\right) \quad (6)$$

where $A$ – gas amplification and $\theta$ - parameter which depends on a working gas and electric field configuration. From the approximation of the measured spectrum by Polya curve it was found that $\theta \approx 0.16$ what is in a reasonable agreement with the expected one for our working gas and electric field configuration. The counting efficiency has been corrected upon the results of calibrations performed in each of three configurations. The possibility to increase the counting efficiency is a subject of our further study. To have a further progress we need to decrease the threshold or to increase the gas amplification.

In the measurements the shapes of the pulses on the output of a charge sensitive preamplifier are recorded by 8-bit digitizer. In our previous paper [9] the shapes of "true" and "wrong" pulses are presented. The "true" pulses have typically a relatively short front edge (a few microseconds) corresponding to the drift of positive ions to cathode and long (hundreds of microseconds) tail corresponding to the time of the baseline restoration of the charge sensitive preamplifier. The "wrong" pulses usually have a wrong (too fast or too slow or irregular) front edge or non-exponential tail. In the analyses of the data only pulses with a baseline within ± 2 mV were taken with a proper evaluation of the resulted live time. Figures 6, 7 show the distribution of the events on the diagram "duration of front edge – parameter β" for the UV lamp and for real measurements. The parameter β is proportional to first derivative of the baseline approximated by a straight line in the interval 50 μs before the front edge. We used it as a measure of the quality of baseline in the prehistory of the event what has been used in a procedure of automatic screening of all data. The events with β beyond the allowed range of 0 ± 0.1 were considered as "not having a reliable determined amplitude" events and were discarded from analyses. The statistic collected during a day or even a few days for each configuration was rather high so we could afford to do this discrimination without substantial loss of information. The region of interest (ROI) box contains "true" pulses with amplitude in the interval [7 – 70] eV with a front edge in the interval [2 - 25 μs] and a parameter β in the interval 0.0 ± 0.1. The pulses beyond this region were rejected as "noisy" pulses. One of typical "noise" pulses is presented on Fig.5.

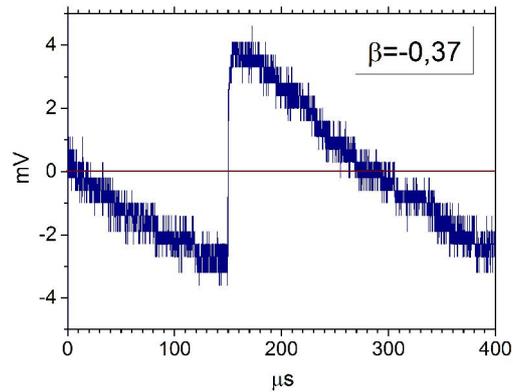

Figure 5. One example of the pulse with "bad" prehistory.

One can see that a ROI region contains 95 ± 5 % of all pulses (crosses) from UV lamp. By inspecting directly a small sample of the real pulses we found that inside of ROI region one can see only about 10% of the pulses with the "wrong" shape (circles).

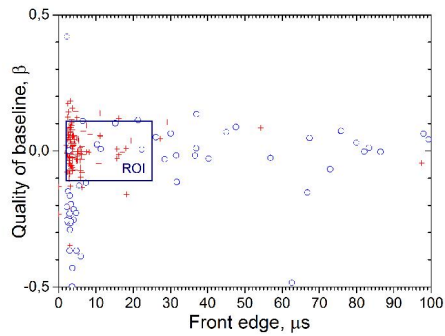

Figure 6. The distribution of the events for UV lamp.

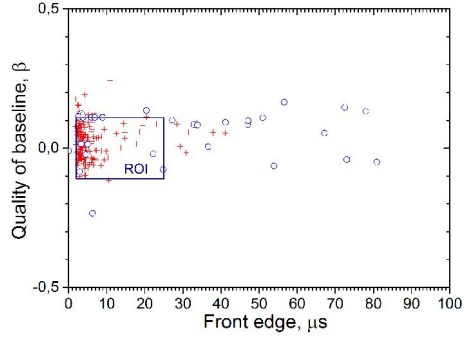

Figure 7. The distribution of the real events.

To reduce the background from external γ-radiation the counter has been placed in a cabinet with 30 cm iron shield. It resulted in decrease of the count rates of single-electron events by a factor of 2 (detector on a porch of a cabinet versus detector inside a cabinet) while the flux of gamma rays in the region around 200 keV has been attenuated in these lay-outs by a factor of 50. From here one gets a simple estimate of the background: inside a cabinet the γ-radiation can't be a major source of the single electron events, its contribution is not more than a few percent. The main source is spurious pulses similar to what one observes with PMT, this was also a limiting factor in experiment with a dish antenna [5]. To reduce this dark current we should make further improvements in the construction of the detector. This is our task for future.

The data were collected frame by frame. Each frame contained 2M points each point 100 ns. After collection of the data they were stored on a disk then the collection resumed. The analysis of the collected data was performed off-line. The frames with the signs of excessive noisiness were removed from analysis.

## 4. Sensitivity of the method

Here we follow the same ideology as developed in [6] for a dish antenna with one principal difference: instead of detecting electromagnetic waves we look for single electrons emitted from the surface of antenna. That is why we use in our experiment not antenna, but gaseous proportional counter (see Fig.1). We assume that similar to the emission of single electrons from metal by UV light or by X-rays the probability for the electron to be emitted when HP of the mass $m_{\gamma'}$ gets converted into an ordinary electric field in metal is equal to the quantum efficiency η for the photon's energy $\omega = m_{\gamma'}$. According to [6] if DM is totally made up of hidden photons, the power collected by antenna is

$$P = 2\alpha^2\chi^2\rho_{CDM}A_{dish} \qquad (7)$$

Where: $\alpha^2 = \cos^2\theta$, θ is the angle between the HP field, when it points in the same direction everywhere, and the plane of antenna, and $\alpha^2 = 2/3$ if HPs have random orientation; χ is the dimensionless parameter quantifying the kinetic mixing, $\rho_{CDM} \approx 0.3$ GeV/cm³ – is the energy density of CDM which is taken here to be equal to the energy density of HPs and $A_{dish}$ – the antenna's surface. In our case of gaseous proportional counter $P = R_{MCC}m_{\gamma'}/\eta$ and this expression will look

$$R_{MCC}m_{\gamma'} = 2\eta\alpha^2\chi^2\rho_{CDM}A_{MCC} \qquad (8)$$

Here $A_{MCC}$ is the surface of the metal cathode of our counter. From here one can easily obtain

$$\chi_{sens} = 2.9 \times 10^{-12} \left(\frac{R_{MCC}}{\eta\, 1\,\text{Hz}}\right)^{1/2} \left(\frac{m_{\gamma'}}{1\,\text{eV}}\right)^{1/2} \left(\frac{0.3\,GeV/cm^3}{\rho_{CDM}}\right)^{1/2} \left(\frac{1\,\text{m}^2}{A_{MCC}}\right)^{1/2} \left(\frac{\sqrt{2/3}}{\alpha}\right) \qquad (9)$$

## 5. First data obtained

The detector was placed at the ground floor of a building in Troitsk, Moscow region in a specially constructed cabinet with 30 cm iron shield. All count rates were in a few Hz range. The average value of $R_{MCC}$ calculated for "quiet" interval during 28 days of measurements was found to be: $\overline{R}_{MCC}$ = - 0.06 ± 0.36 Hz. The uncertainty has been found from the real scattering of the experimental points. So if we take the normal distribution for uncertainties, then we obtain that at 95% confidence level: $\overline{R}_{MCC}$ < 0.66 Hz. The quantum efficiency η was taken from [17] for masses of HPs $m_{\gamma'}$<11.6 eV (magenta), from [18] for 10 eV < $m_{\gamma'}$< 60 eV (red), from [19] for 20 eV < $m_{\gamma'}$ < 10 keV (green) and from [20] for 50 eV < $m_{\gamma'}$< 10 keV (blue). From the expression (9) we obtain an upper limit for a mixing constant χ. The values of a mixing constant χ allowed by this experiment are below the curve presented on Figure 8. The systematic uncertainty is mainly determined by the uncertainty in quantum efficiency which is taken to be about 30% following the estimates done in [17]. To decrease this limit one should construct a detector with lower count rate of spurious pulses. The difference in the curves presented at Fig.8 is explained by different purity of copper used in measurements. The data from [17] and [20] were obtained for atomically clean copper samples prepared by evaporation of copper in high vacuum while routinely cleaned (by solvents) copper samples were used in [18, 19]. For example in the paper [21] it was shown that cleaning of the surface of the copper cathode by ionized controlled etching (ICE) can increase the quantum efficiency by an order of magnitude. For atomically clean copper one can see the effect of electronic shells while for routinely cleaned copper the spectra are rather smudged. In our detector for cathode we used routinely cleaned (by solvents) copper. It would be expedient to use a quantum efficiency measured for the specific sample used for a cathode in our detector and we are planning this work for the future. There is a strong dependence of the effect not only on the work function but also on the structure of the electronic shells of the metal used for a cathode. Potentially this can be used for verification that the obtained result is really from HPs. For this we should make measurements using cathodes made of different metals.

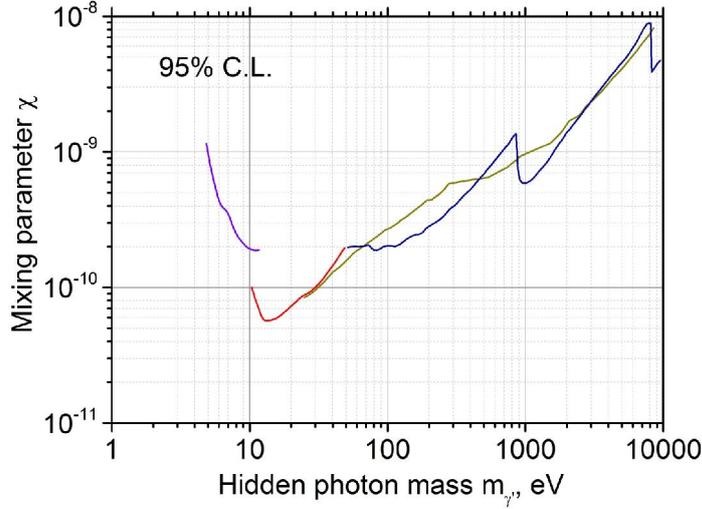

Figure 8. The limit for a mixing constant χ.

The numbers presented at Fig.8 are first, very preliminary results for hidden photons with a mass from 5 to 500 eV. One can see that the region of highest sensitivity for our method is from 10 to 30 eV i.e. approximately equal to the energy needed to produce one ion pair in argon. This result has been obtained in direct measurements in a laboratory experiment by observing single electrons emitted from the surface of a copper cathode. At present time this is the only experiment using this technique which is based on measurement of the rate of single electron emission from a metal cathode of the counter sensitive only to HP-photon conversion and not sensitive to photon-HP conversion. Stellar astrophysics provides stringent constraints for this value. The limits obtained by using some astrophysical models, see for example [22, 23] and references therein, are lower by several orders of magnitude. The limits obtained from stellar astrophysics are based on the models when both conversions are alike. The most impressive results have been obtained by observing electron emission from liquid xenon [23]. The threshold energy for the production of a single electron in liquid xenon is only 12 eV and in a xenon detector the total fiducial volume is a target for CDM. To reach the comparable sensitivity we need to decrease the spurious count rate of single electron events of our detector by several orders of magnitude. We are not aware of any reasons why it can't be realized in view that this is the first multi-cathode counter ever constructed. This question needs a further study. We plan to make further improvements in the construction of MCC mainly with the aim to reduce the count rate of spurious pulses.

We consider that a key element in reducing the rate of spurious counts would be to use good isolating materials, very clean metals with meticulously polished surfaces and to assemble the counter in very clean dust free environments. Approaching this strategy we hope to decrease the rate of spurious counts by about an order of magnitude. At this level, as our measurements with and without an active and passive shields have shown, the background rate will be determined by ionized particles (muons and electrons) and by gammas from the material of passive shield. To decrease further the background rate we will go to underground laboratory, where the flux of muons is negligible in comparison with the one at a surface laboratory, and there we will also use a passive shield cabinet similar to the one we use now at the surface laboratory in Troitsk,

Moscow region. Obviously the question of how far one can progress on this way can be solved only experimentally.

**6. Conclusion**

A new technique of Multi Cathode Counter (MCC) has been developed to search for hidden photon CDM in the assumption that all dark matter is composed of hidden photons (HP). It was assumed also that if HPs have a mass greater than a work function of the metal of a cathode, they will induce emission of single electrons from a cathode. The technique used in this experiment is sensitive to HP – photon conversion and is not sensitive to photon – HP conversion. First preliminary result has been obtained for HPs with a mass from 5 to 500 eV which demonstrate that this method works. The upper limit is above the ones obtained from stellar astrophysics by two - three orders of magnitude (see Fig.9 taken from [24]), but our result (MCC, yellow lines) has been obtained in direct measurements in a laboratory experiment. It is worth also to note that our result has been obtained for the conversion of HPs of CDM into usual photon while the result obtained in astrophysics has been obtained for the conversion of photons in solar plasma in the depth of the Sun into HPs, what is impossible to test experimentally.

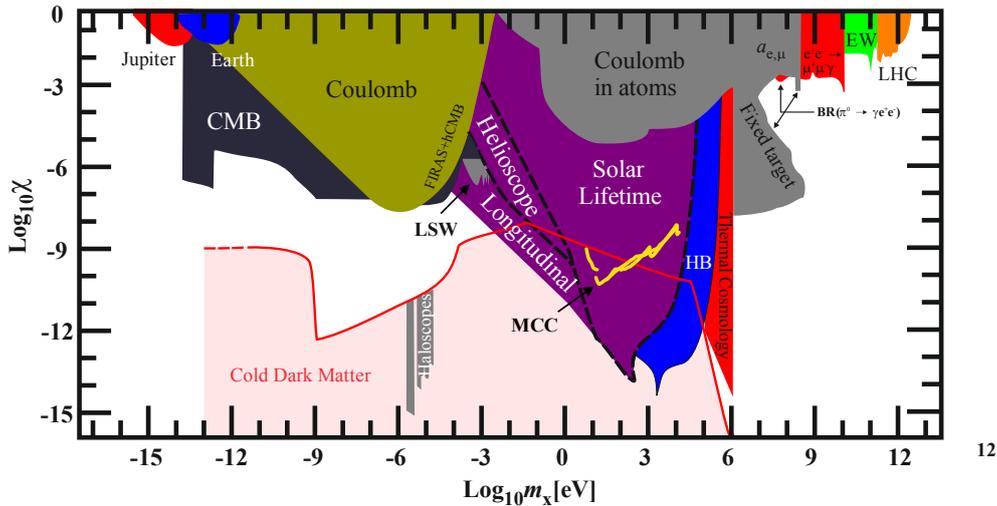

Figure 9. Constrained and still testable regions of photon/hidden-photon mixing.

Recently we are working with the aim to decrease our limit substantially. Further progress will depend upon how successful will be our efforts to construct a detector with a lower count rate of spurious pulses. If they are successful and all sources of spurious pulses are eliminated then the only source will be left, which can be eliminated by no means, it's a dark matter. The same will be true also for PMTs provided that dark matter is really composed of hidden photons and their mass is greater than a work function of a metal used for internal elements of detector. At present time we are constructing a new detector with a more developed design.

**Acknowledgements**

The authors express deep gratitude to E.P.Petrov and A.I Egorov for very substantial contribution to fabrication of the counter and to Grant of Russian Government "Leading Scientific Schools of Russia" #3110.2014.2 for partial support of this work.

**Conflict of Interests**

The author declares that there is no conflict of interests regarding the publication of this paper.